# FPGA Based Pico-second Time Measurement System for a DIRC-like TOF Detector

Qiang Cao, Xin Li, Liwei Wang, Jie Kuang, Yonggang Wang and Cheng Li

*Abstract*—A prototype of DIRC-like Time-of-Flight detector (DTOF), including a pico-second time measurement electronics, is developed and tested preliminarily. The basic structure of DTOF is composed of a fused silica radiator connected to fast micro-channel plate PMTs (MCP-PMT), and readout by a dedicated FPGA (Field Programmable Gate Array) based front-end electronics. The full electronics chain consists of a programmable differential amplifier, a dual-threshold differential discriminator, and a timestamp Time-to-Digital convertor. By splitting a MCP-PMT output signal into two identical electronics chains, the coincidence time resolution (CTR) of pure electronics was measured as 5.6 ps. By the beam test in H4 (150GeV/c, Muon) at CERN, the intrinsic CTR of the whole detector prototype reaches 15.0 ps without using time-amplitude correction. The test results demonstrate that the FPGA based front-end electronics could achieve an excellent time performance for TOF detectors. It is very compact, cost effective with a high multi-channel capacity and short measurement dead time, which is very suitable for practical applications of large-scale high performance TOF detectors in particle physics spectrometer.

## I. Introduction

DIRC-like Time-of-Flight detector (DTOF) is an innovative TOF utilizing internally reflected Cherenkov light for high energy charged particle identification [1]. It achieves a high level of performance at the extreme data taking conditions under high luminosity and high backgrounds. The basic structure of DTOF is composed of a fused silica radiator connected to fast photomultiplier (MCP-PMT or SiPM) array, readout by dedicated front-end electronics. To construct such a detector, the challenge comes from the fact that the limited flight length for charged particles in detectors requires the time measurement, including all uncertainties in the detector, at the tens of pico-second level. For example, the High Luminosity of the Large Hadron Collider (HL-LHC) at CERN is expected to provide instantaneous luminosities of $10^{34}$ cm$^{-2}$s$^{-1}$ and above. Since the size of the detector is a few meters, to associate EM calorimeter measurements with primary vertex location, it requires TOF's time resolution less than 30ps to suppress pileup in collisions.

The time uncertainty contributed from the front-end electronics is expected as low as possible. Furthermore, it is desired to be compact and cost-effective for future integration of multiple channels. Besides of an amplifier required for conditioning the input signal in the electronics chain, the timing mechanism and related circuit realization should be the most important part. Normally there are two basic timing options: 1) employ a constant-fraction discriminator (CFD) followed by a time-to-digital convertor (TDC), or 2) determine the time by waveform sampling, storing, and analyzing the full signal shape [2]. Although CFD technique has small time-walk for different signal amplitudes, the realization circuit is relatively complex, which is not feasible for future multi-channel integration. While the second option is a promising approach for low rate application, it is much more challenging for high flux use and large-scale multi-channel integration. In this paper, we propose a new dual-threshold front-edge timing scheme. Not only can high timing performance be achieved [3], but also the discriminator can be realized using commercial field programmable gate array (FPGA) chip. The followed TDC is also implemented inside of FPGA, so that the electronics have very high integration [4]. The whole detector prototype was developed and its performance was preliminarily tested with β-ray in our lab and specific beam at CERN. By splitting a MCP-PMT signal generated by β-ray, the coincidence time resolution (CTR) of two identical electronics chains was measured as 5.6 ps, meanwhile, the beam test show that the CTR of the two detectors reaches 15.0 ps. Before a conclusion is drawn at the end of this paper, the design of the prototype and the performance test process will be introduced in the following sections.

## II. Pico-second Time Measurement System

Fig. 1(a) shows the structure diagram of pico-second time measurement system, which mainly consists of a pre-amplifier, a dual-threshold differential discriminator (DDD), and a Time-to-Digital convertor (TDC). Except the amplifier and the resister-capacitor bias network, other components are implemented inside of an FPGA (Kintex-7 from Xilinx).

For high time performance, it is crucial that pre-amplifier has a high bandwidth and large measurement dynamic range. In this system, the programmable differential amplifier (PDA) LMH6882 from Texas Instruments was chosen to amplify the

Manuscript received February 2 2018. This work was supported in part by the National Natural Science Foundation of China (NSFC) under Grants 11475168 and 11735013.

Authors are with the Department of Modern Physics, University of Science and Technology of China, Hefei, Anhui, 230026, China.

Corresponding authors: Xin Li (email: li124ste@ustc.edu.cn) and Yonggang Wang (email: wangyg@ustc.edu.cn).

PMT signal. The bandwidth of this amplifier is 2.4 GHz over the gain range from 6 dB to 26 dB. Furthermore, this amplifier provides conversion from single-end input to differential output, which transfers the PMT signal to the DDD module with a strong anti-noise ability.

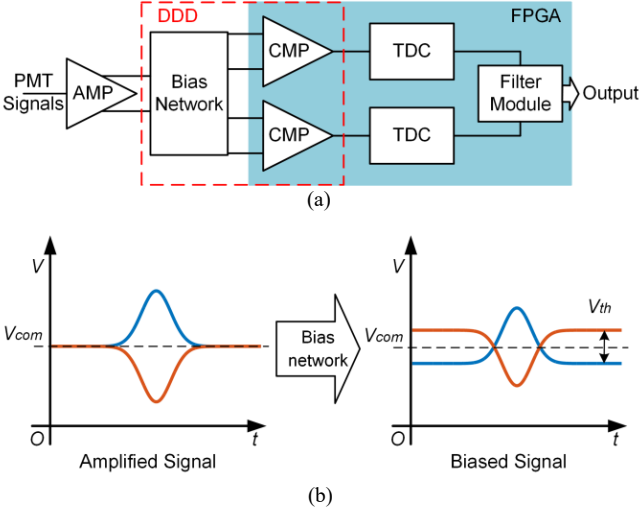

Fig. 1. (a) Schematic diagram of pico-second time measurement system. (b) Sketch map of biasing process.

DDD module is composed of a resistor-capacitor bias network and two high performance comparators based on LVDS receivers of FPGA. The amplified differential signals are first biased by the bias network and translated as the form suitable for LVDS comparator. The positive and negative signals are biased on two different offset voltage levels with their common-mode voltage (marked as $V_{com}$ in Fig. 1(b)) unchanged. The difference of the two offset voltages is the threshold voltage marked as $V_{th}$ in Fig. 1(b). In the DDD module, the input signals are split and fed into two differential discriminators, each of them has different thresholds. Both discriminators are front-edge timing. The output signal from the comparator with a low threshold will be the event timing while the one with higher threshold is used as a confirming signal. The dual-threshold mode has small timing walk for input signals with different amplitudes. The outputs of these two comparators are fed into two channels of TDC as shown in Fig. 1(a).

Our TDC module is implemented in the FPGA using ones counter encoding scheme, which was presented in our previous work [3]. The RMS time resolution was evaluated as 3.9 ps in our previous work. At the end of electronics chain, there is a filter module for event judgement. The measured event timing of good events are then transmitted to PC for further processing.

### III. DIRC-LIKE TOF DETECTOR

Our homemade DIRC-like TOF detector is composed of a MCP-PMT (R3809U from Hamamatsu) connecting to a 15 mm × 15 mm × 20 mm square radiator (JC-H02 fused silica from Beijing Quartz and Special Glasses Institute), which has more than 90% transmission rate for ultraviolet with wavelength about 200nm. All surfaces of the radiator are polished and wrapped with black lightproof tape except the surface facing to the MCP-PMT. According to the datasheet of R3809U and our test result, the rise time of signal from the TOF detector is about 160 ps.

### IV. PRELIMINARY TEST RESULTS AND DISCUSSION

#### A. Electronics Performance Test

We built an experiment setup to test the performance of the pico-second time measurement system using β-ray in our lab. To observe the time jitter contributed by pure electronics, the anode signal from one DIRC-like TOF detector was split for two identical electronics chains. For each event, both measured timestamps from the two electronics chains were read out to the PC, and the time difference between them was calculated. The diagram of the experimental setup is shown in Fig. 2(a), where the CM represents a coincidence module added for the test. The measured CTR of pure electronics chains is shown in Fig. 2(b). With Gaussian fitting, the standard deviation, i.e. the RMS CTR of two identical electronics chains, is calculated as 5.6 ps.

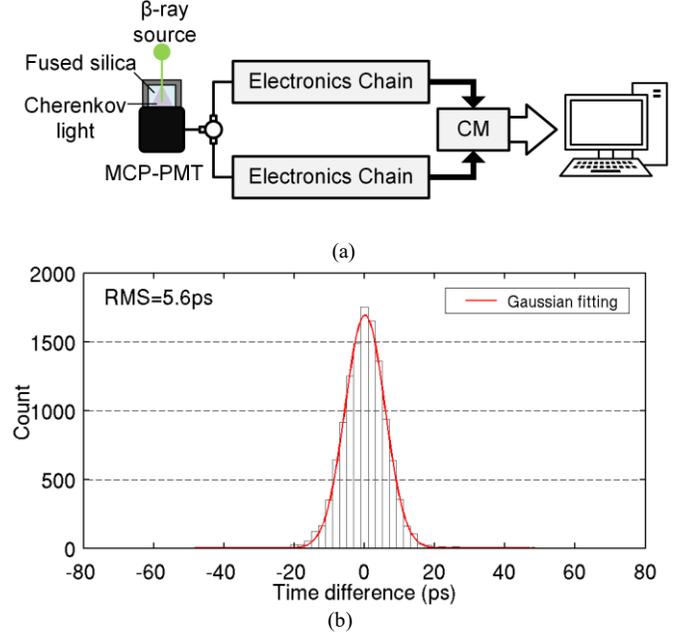

Fig. 2. (a) Schematic of electronics performance test experimental setup. (b) Histogram of time difference measured in electronics performance test experiment.

#### B. Beam Test Result

The performance of TOF detector, along with the electronics system, was evaluated by specific test beam in H4 (150GeV/c, Muon) at CERN. The diagram of experimental setup is shown in Fig. 3(a). Fig. 3(b) shows that two identical detectors were fixed and aligned by a metallic holder to detect the test beam simultaneously. The anode signals came out from detector when charged particles transmitting through two detectors. For each detector, signals were fed into one of electronics chains implemented on our circuit board. After event selection based on coincidence module implemented in FPGA, the measured timestamps of detected events were led out to the host computer. The time difference was then calculated offline. The CTR of these two TOF detectors are measured as shown in Fig. 3(c), where the RMS error is 15.0 ps.

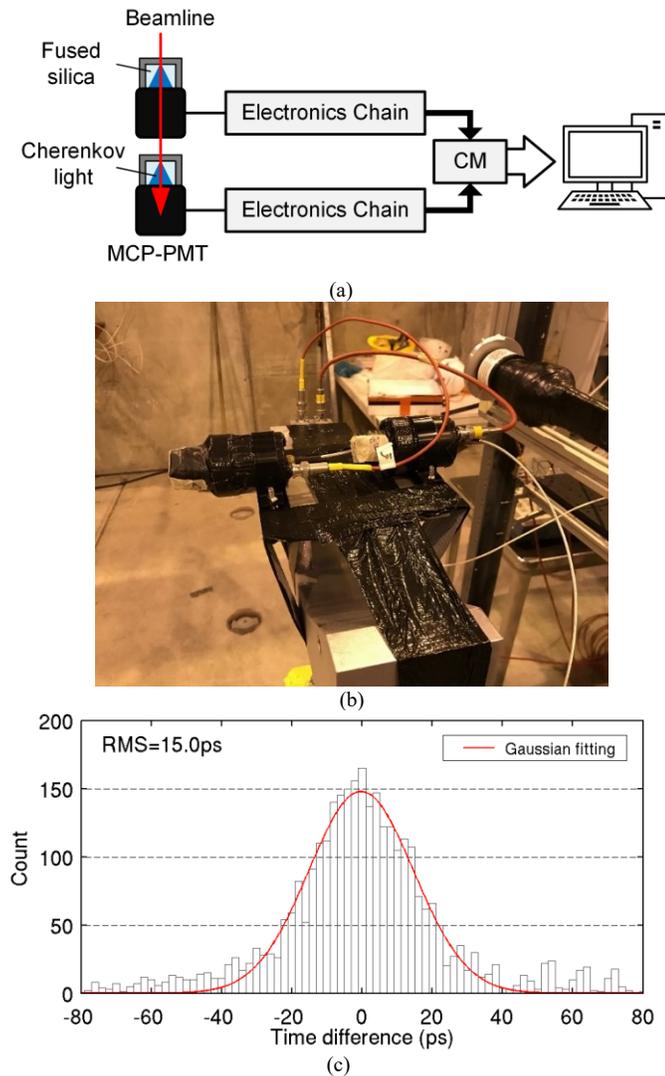

Fig. 3. (a) Schematic of beam test experimental setup. (b) The experiment photograph taken at CERN. (c) Histogram of time difference measured in beam test experiment.

Because the result of Fig. 3(c) was obtained using a wide test beam, the transmitting time of scattering light inside of radiator has a significant contribution to the test result. Furthermore, the dual threshold values and the detector performance were not optimized. Therefore, the test result of detectors is very preliminary. It is reasonable to believe that the detector performance will be better when these two aspects are handled properly.

## V. Conclusion

Pico-second time measurement electronics system is a very challenging part in TOF detectors, in particular for high flux collider and multi-channel applications. The electronics scheme proposed in this paper can minimize the number of components, while providing an excellent time performance, which shows great potential in applications for TOF detectors. Considering the dual threshold values of the discriminator were not optimized during the beam test, and also the beam was not a fine one for timing test, it is reasonable to believe that the tested CTR value is not its optimum performance. It is estimated that the CTR value can be less than 10 ps when the two factors are properly handled. The revised electronics along with the detector will be finally tested in this July.